\documentclass[reprint,superscriptaddress]{revtex4-2}
\usepackage[T1]{fontenc}
\usepackage[utf8]{inputenc}
\usepackage{url}
\usepackage{amsmath}
\usepackage{amssymb}
\usepackage{mathtools}
\usepackage{bbold}
\usepackage{mathrsfs}

\usepackage{graphicx}
\usepackage{epstopdf} 
\usepackage{wrapfig}
\usepackage{braket}
\usepackage{xcolor}
\usepackage{float}
\usepackage{mathptmx}
\DeclareMathAlphabet{\mathcal}{OMS}{cmsy}{m}{n}
\usepackage{tikz}
\usepackage[vcentermath]{youngtab}
\usepackage{cancel}
\begin{document}

\newcommand{\blu}[1]{{\color{blue}#1}}

\author{Gianni Aupetit-Diallo}
\affiliation{Université Côte d’Azur, CNRS, Institut de Physique de Nice, 06560 Valbonne, France}
\author{Giovanni Pecci}
\affiliation{Univ. Grenoble Alpes, CNRS, LPMMC, 38000 Grenoble, France}
\author{Charlotte Pignol}
\affiliation{Université Côte d’Azur, CNRS, Institut de Physique de Nice, 06560 Valbonne, France}
\author{Fr\'ed\'eric H\'ebert}
\affiliation{Université Côte d’Azur, CNRS, Institut de Physique de Nice, 06560 Valbonne, France}
\author{Anna Minguzzi}
\affiliation{Univ. Grenoble Alpes, CNRS, LPMMC, 38000 Grenoble, France}
\author{Mathias Albert}
\affiliation{Université Côte d’Azur, CNRS, Institut de Physique de Nice, 06560 Valbonne, France}
\author{Patrizia Vignolo}
\affiliation{Université Côte d’Azur, CNRS, Institut de Physique de Nice, 06560 Valbonne, France}

\title{Exact solution for $SU(2)$-symmetry breaking bosonic mixtures at strong interactions}

\begin{abstract}
We study the equilibrium properties of a one-dimensional mixture of two Tonks-Girardeau gases on a ring geometry in the limit of strongly-repulsive inter-species interactions. We derive the exact many-body wavefunction and compare it to the $SU(2)$ solution where intra- and inter-species interactions are also diverging but equal.  We focus on the role of the $SU(2)$-symmetry breaking on the behaviour of the large- and short-distance correlations by studying the zero-momentum occupation number and the Tan's contact from the asymptotic behavior of the momentum distribution.
Although the symmetry is only weakly broken, it has important consequences on spin correlations in the system 
as the reduction by a factor of two of the zero-momentum occupation number with respect to the $SU(2)$ case in the thermodynamic limit and the decrease of the Tan's contact.
 \end{abstract}

\maketitle
\section{Introduction}
Ultracold atomic mixtures are an important paradigm for quantum simulators due to their extreme versatility. Such systems offer the possibility to control most of the microscopic parameters such as dimensionality, interaction strength and range, the number of spin components, the number of atoms and  external potentials while giving access to many physical observables, including those intimately connected to quantum correlations.

In particular one-dimensional (1D) mixtures are of significant importance \cite{Yurovsky2008,cazalilla2011onedimensional,guan2013fermi,Sowinski2019,Minguzzi2022,Mistakidis2022}
since correlations are enhanced by the reduced dimensionality. They also offer the advantage to access, in some special cases, the exact many-body wavefunction. For instance, homogeneous 1D quantum systems with a well-defined symmetry can be solved via Bethe Ansatz \cite{Gir1965,Lieb,Yang67,Sutherland68,mcguire1964study,calabrese2007correlation,piroli2016local,Gaudin67,guan2011quantum}). However, it is well known that correlation functions often remain very difficult to extract due to the complexity of the Bethe Ansatz equations and the resulting many-body wavefunction.

One special case that allows to go further in the calculations, capturing correlation functions \cite{Deuretzbacher,Decamp2016,Decamp2016-2,Decamp2017} and dynamics \cite{Deuretzbacher2014,Volosniev2015,Deuretzbacher2016,pecci2021universal}, even in the presence of external confinements, is the Tonks-Girardeau (TG) limit \cite{Gir1965}, where the interaction strength is repulsive and tends to infinity. The study of this limiting case allows a deep understanding of quantum correlations in many-body systems \cite{Rizzi2018,Santana2019,Capuzzi2020} in and out-of equilibrium, and to have a benchmark for numerical simulations of such systems.

The ground state of such TG
quantum mixtures is highly degenerate due to exchange symmetry of particles. Indeed, at zero temperature, any arrangement of the particles has the same energy. On the other hand, in an actual experiments, interaction between particles although potentially large, always remain finite and consequently the macroscopic degeneracy is lifted and the lowest-energy state is generally unique and corresponds, for the spatial part, to the most symmetric possible state \cite{LiebMattisPR,Decamp2016}.

In this article we show that the ground state for a 1D strongly interacting 
mixture depends on the protocol used to approach 
the TG regime.
We consider a two-component bosonic mixture and we analyse two cases: the $SU(2)$ case where the intra- and inter-species interactions are equal and very large, and the symmetry breaking case (SB) where the intra-species interactions are diverging and the inter-species interaction is increased afterwards.  
In the first case the many-body ground state is identical to that of a single component TG gas: the spatial symmetry is the highest
and the two spin components are strongly correlated even at large distance because, due to the symmetry, it is as if there was no spin at all.
In the SB case, we show that the distinguishability introduced by the difference between the inter- and intra-species interaction strengths, even if it slightly affects the symmetry of the many-body wavefunction, makes the spin correlation to drastically drop.

The manuscript is organized as follows. The model is presented in Sec. \ref{sec-model}. In this section we discuss in detail our procedure to obtain the ground-state many-body wavefuntion for the $SU(2)$ Hamiltonian and for the SB one. We quantify the breaking of the symmetry associated to the SB many-body ground-state by calculating the expectation value of the 2-cycle-sum operator in Sec. \ref{sec-sym}.
We show that for large number of particles, the SB state is halfway between the most symmetric and the most anti-symmetric states
allowed by the $SU(2)$ Hamiltonian. Correlations are analyzed starting from Sec. \ref{sec-corr}. We calculate the momentum distribution that is given by the Fourier Transform of the one-body density matrix. In Secs.~\ref{sec-n0} and \ref{sec-tan}  we study the zero-mode occupation number and the Tan's contact. The first is related to long-distance correlations and the second to short distance correlations.
In Sec. \ref{sec-concl}, some remarks on the relation between our approach and the Bethe ansatz solution conclude the paper.
\section{The model}
\label{sec-model}
We consider a balanced two-components 1D Bose gas, characterized by contact interactions, in a ring geometry (periodic boundary conditions) at zero temperature. The two components are labeled by an index $\sigma=\uparrow,\downarrow$ for convenience but this index has no relation with the spin one-half of fermions. The general Hamiltonian for $N$ bosons reads
\begin{equation}
\begin{split}
\hat{H}&=\sum_{\sigma=\uparrow,\downarrow}\sum_i^{N_\sigma}
\left[-\frac{\hbar^2}{2m}\frac{\partial^2}{\partial x_{i,\sigma}^2}+g_{\sigma\sigma}\sum_{j>i}^{N_{\sigma}} \delta(x_{i,\sigma}-x_{j,\sigma})\right] \\
&+g_{\uparrow\downarrow}\sum_{i}^{N_\uparrow}\sum_{j}^{N_\downarrow}\delta(x_{i,\uparrow}-x_{j,\downarrow})
\end{split}
\label{ham}
\end{equation}
with $g_{\uparrow\downarrow}$ the inter-species, $g_{\uparrow\uparrow}$ ($g_{\downarrow\downarrow}$) the intra-species interaction strengths and $N_\downarrow=N_\uparrow=N/2$ the number of particle per component. The aim of this work is to analyse the ground-state coherence properties of the symmetry breaking case with $g_{\uparrow\uparrow}=g_{\downarrow\downarrow}\neq g_{\uparrow\downarrow}$ with respect to
those of the $SU(2)$ case with $g_{\uparrow\uparrow}=g_{\downarrow\downarrow}=g_{\uparrow\downarrow}$.
 
The contact interactions can be accounted for by the cusp conditions on the many-body wavefunction, which read
\begin{equation}
    \left[\dfrac{\partial\Psi}{\partial x_{\sigma,i}}-\dfrac{\partial\Psi}{\partial x_{\sigma,i'}}\right]^{x_{\sigma,i}-x_{\sigma,i'}=0^+}_{x_{\sigma,i}-x_{\sigma,i'}=0^-}=\dfrac{2mg_{\sigma\sigma}}{\hbar^2}\Psi(x_{\sigma,i}=x_{\sigma,i'}),
    \label{cuspC}
\end{equation}
\begin{equation}
    \left[\dfrac{\partial\Psi}{\partial x_{\uparrow,i}}-\dfrac{\partial\Psi}{\partial x_{\downarrow,i'}}\right]^{x_{\uparrow,i}-x_{\downarrow,i'}=0^+}_{x_{\uparrow,i}-x_{\downarrow,i'}=0^-}=\dfrac{2mg_{\uparrow\downarrow}}{\hbar^2}\Psi(x_{\uparrow,i}=x_{\downarrow,i'}).
    \label{cuspC2}
\end{equation}

 \subsection{The ground-state solution in the strongly interacting limit}

In the limit $g_{\sigma\sigma'}\rightarrow +\infty$, for any $\sigma,\sigma'$, the many-body wavefunction vanishes whenever $x_i=x_j$
(from now on we drop the spin index in the particle positions). Thus, it can be written in terms of linear combinations of fermionic wave functions \cite{Volosniev2014,Deuretzbacher2014}
\begin{equation}
\Psi(x_1,\dots,x_N)=\sum_{P\in S_N}a_P\theta_P(x_1,\dots,x_N)\Psi_S(x_1,\dots,x_N)
\label{vol}
\end{equation}
 where $S_N$ is the permutation group of $N$ elements, $P$ a permutation operator, $\theta_P(x_1,\dots,x_N)$ is
equal to 1 in the coordinate sector $x_{P(1)}<\dots<x_{P(N)}$. The wavefunction $\Psi_S=A\Psi_A$ is given by the action of the unit antisymmetric function $A=\prod_{i<j}{\rm sgn}(x_i-x_j)$  on
 the fully antisymmetric fermionic wavefunction,
\begin{equation}
    \Psi_A=\dfrac{1}{\sqrt{N!}}\det\left[\phi_m(x_n)\right],
\end{equation}
where $\phi_m(x_n)=e^{ik_m x_n}/\sqrt{L}
    $    with $k_m=\pi(2m-N-1)/L$ and $n,m\in1,\dots,N$, $\forall x_n\in[-L/2,L/2]$, for particles on a ring of length $L$.
The rules for exchanging identical particles being fixed by the statistics, we can restrict our basis to $N!/(\frac{N}{2}!\frac{N}{2}!)$
 independent sectors (and then $a_P$'s) instead of the $N!$ possible.
These sectors represent all the possible spins configurations and are usually called snippets \cite{Deuretzbacher}. They constitute the proper basis to describe a two-component spin mixture and will be used all along this manuscript.

In this work we focus on the ground state solution that is not degenerate for balanced mixtures (in the limit $g_{\sigma\sigma'}$ very large but finite), so that we can set $a_P$ real 
without loss of generality,
and use a strong-coupling expansion
approach \cite{Volosniev2013},
by calculating the energy
to first order with respect to the small parameters $1/g_{\sigma\sigma'}$.

\subsubsection{The $SU(2)$ case}
We first recall the method for the 
case of a $SU(2)$ boson gas with $g_{\uparrow\downarrow}=g_{\uparrow\uparrow}=g_{\downarrow\downarrow}=g$. The two-spin components being indistinguishable, we expect the ground state to be equivalent to that of a one-component TG gas. In such a case all the $a_P$'s are equal in each sector. 

The minimization of the energy in the limit
$g\rightarrow\infty$, 
\begin{equation}
E_g\simeq E_{\infty}+\dfrac{1}{g}[\partial_{1/g}E]_{g\rightarrow\infty}=E_{\infty}-\dfrac{1}{g}K
\label{eq-min}
\end{equation}
corresponds to the maximization of the energy slope $K=-[\partial_{1/g}E]_{g\rightarrow\infty}$.
The procedure is analogous to that outlined in \cite{Decamp2016}. One writes $K$ as a function of the $a_P$ coefficients,
\begin{equation}
K(a_P)=\dfrac{\hbar^4}{m^2}\left(\sum_{P,Q\in S_N}(a_P+a_Q)^2\alpha_{P,Q}+2\sum_{P,P'\in S_N}a_{P'}^2\alpha_{P,P'}\right),
\label{eqK}
\end{equation}
and then finds the stationary solutions of this function
taking into account the normalization condition $\sum_P a_P^2=1$.
The terms $\alpha_{P,Q}$ in Eq. (\ref{eqK}) are the nearest-neighbour exchange constants, given by the relation
\begin{eqnarray}
\alpha_{P,Q}&=&N!\int{\rm d}x_1,\!\dots\! {\rm d}x_N \theta_{\rm Id}(x_1,\dots,x_N)\delta(x_k\!-\!x_{k+1})
\left[\dfrac{\partial \Psi_A}{\partial x_k}\right]^2\nonumber\\
&\equiv& \alpha_k \label{Eq:alpha}
\end{eqnarray}
if $P$ and $Q$ (i.e. $P$ and $P'$) are equal up to a transposition
of two consecutive distinguishable particles (i.e. indistinguishable bosons), and $\theta_{\rm Id}(x_1,\dots,x_N)$ is the indicator of the sector $x_1<\dots<x_N$.

In a ring geometry, at fixed number of particle $N$, all the $\alpha_k$'s are the same, $\alpha_k=\alpha^{(N)}$, $\forall k$, because of the homogeneity of the potential. In order to calculate $\alpha^{(N)}$, we consider one of the variable (here $x_1$) as a moving boundary for the other $N-1$ ones. This leads to
\begin{eqnarray}
    \alpha^{(N)}&=&(N-1)!\int_0^Ldx_1\prod_{i=2}^{N}\int_{x_{i-1}}^{x_1+L}dx_{i}\delta(x_1-x_{2})\left|\frac{\partial\Psi_A}{\partial x_1}\right|^2\nonumber\\
    &=&\frac{N(N^2-1)}{3L^3}\pi^2.
\end{eqnarray}
In agreement with \cite{Barfknecht2021}, we find that $\alpha^{(N)}$ is equal to twice the kinetic energy up to a dimensional constant, the sum of $k_m^2L^2$ over the occupied orbitals being equal to $N(N^2-1)\pi^2/3$. 

The conditioned maximization of $K(a_P)$ is equivalent to solving the eigenvalue problem for a matrix $V$ whose form depends on the type of mixture and trapping potential.
In the bosonic $SU(2)$ case,
\begin{align}
   [V^{SU}]_{i,j}= \frac{\hbar^4}{m^2}\left\{\begin{array}{cc}\sum_{d,k\neq i}\alpha_{i_k}+2\sum_{b,k\neq i}\alpha_{i_k} & j=i \\
   \alpha_{i,j}& j\neq i \end{array}\right.
   \label{vsu}
\end{align}
where the $d$-sum has to be taken over snippets $k$ that transpose distinguishable particles, while the $b$-sum runs over sectors that transpose identical bosons. The explicit form of $V^{SU}$ for the case of a mixture of 2+2 bosons is given in Appendix \ref{app-A}. We remark that the positive sign of the off-diagonal part as well as the plus in Eq. (\ref{eqK}) depend on the choice made while building the many-body wavefunction in Eq. (\ref{vol}), namely, on the choice to start with $\Psi_S$ or $\Psi_A$.

The largest eigenvalue of this matrix, 
\begin{equation}
    K^{SU}=[\Vec{a}_P^{SU}]^t V^{SU}\Vec{a}_P^{SU},
\end{equation}
$\Vec{a}_P^{SU}$ being the eigenvector of $V^{SU}$ corresponding to this eigenvalue,
can be written under the form 
$K^{SU}=K_{\uparrow\downarrow}^{SU}+\sum_{\sigma=(\uparrow,\downarrow)}K_{\sigma\sigma}^{SU}$, %enlightening
highlighting
the inter-component and intra-component contributions to the energy. Finally, we obtain $K^{SU}=2N\alpha^{(N)}\hbar^4/m^2$.

\subsubsection{The symmetry breaking case}

We now move to the more complicated case of two interacting TG gases, where $g_{\uparrow,\uparrow}$ and $g_{\downarrow,\downarrow}$ are infinite and the inter-components interaction strength $g_{\uparrow,\downarrow}$ is very large, but finite.
The minimization procedure, outlined in Eq. (\ref{eq-min}), with respect to the small parameter $1/g_{\uparrow,\downarrow}$
leads to a matrix $V^{SB}$ that does not take into account any intra-component interaction terms:
\begin{align}
   [V^{SB}]_{i,j}= \frac{\hbar^4}{m^2}\left\{\begin{array}{cc}\sum_{d,k\neq i}\alpha_{i_k}&  j=i \\
   \alpha_{i,j}&j\neq i\end{array}\right..
\end{align}
Remark that the largest eigenvalue of $V^{SB}$, denoted as 
$K^{SB}_{\uparrow\downarrow}$
as well as the other eigenvalues, gives only an inter-component contribution to the energy, as $g_{\sigma\sigma}$ has been sent to infinity from the beginning. Of course
the symmetry breaking occurs for $N>2$, as no intra-interaction occurs for the case $N_\uparrow=N_\downarrow=1$.
Again, the explicit form of $V^{SB}$ for the case of a mixture of 2+2 bosons is given in Appendix \ref{app-A}.

We note that $V^{SB}$ is very similar to the matrix $V^{SU}_F$ for a $SU(2)$ fermionic mixture. Indeed $[V^{SU}_F]_{i,i}=[V^{SB}]_{i,i}$,
and $[V^{SU}_F]_{i,j}=-[V^{SB}]_{i,j}$ if $i\neq j$.
The two matrices have the same eigenvalues but the eigenstates do not have the same symmetry, which is well defined for
the case of $SU(2)$ fermions but is not, as we will see in the next section, for the case of two interacting TG gases. Let us point out  that, because of our basis choice, $V^{SU}$ and $V^{SU}_F$ can be mapped on a XXX spin-chain model \cite{Deuretzbacher2016}, while $V^{SB}$ can be mapped on a XXZ model (Appendix \ref{app-xxx}).

\section{Analysis of the symmetry breaking}
\label{sec-sym}
We now explain how to characterize the symmetry properties of the two different ground states using irreducible representations of the permutation group $S_N$. We will show in this section that the ground state of the $SU(2)$ hamiltonian has a well defined symmetry whereas the one of the symmetry breaking case does not. In  order  to quantify the  symmetry  breaking associated  to the many-body state
\begin{equation}
\Psi^{SB}(x_1,\dots,x_N)=\sum_{P\in S_N}a_P^{SB}\theta_P(x_1,\dots,x_N)\Psi_S(x_1,\dots,x_N),    
\end{equation}
we calculate the expectation value of the 2-cycle class-sum operator $\Gamma^{(2)}=\sum_{i<j}(i,j)$ \cite{Kerber,Liebeck},  whose  eigenvalues  are  directly connected to the irreducible representations of $S_N$, and thus to the Young tableaux. 
Indeed the relation between the eigenvalues $\gamma^{(2)}$'s and a Young tableaux with a number of boxes $\lambda_i$ at line $i$ is
\begin{equation}
\gamma^{(2)}=\dfrac{1}{2}\sum_i[\lambda_i(\lambda_i-2i+1)].
\end{equation}
Thus, for the fully symmetric $SU(2)$ ground state, corresponding to the Young tableau $(N)={\tiny\yng(3)}\cdots{\tiny\yng(3)}$, one has $\gamma^{(2)}_S=N(N-1)/2$, namely $\gamma^{(2)}_S$ is given by the number of pairs in a system of $N$ particles.
Instead the antisymmetric eigenvalue $\gamma^{(2)}_A$, corresponding to the Young tableau
$(N/2,N/2)={\tiny\yng(3,3)}\cdots{\tiny\yng(3,3)}$, is equal to $N(N-4)/4$. This corresponds to the number of pairs in a system of $N/2$ particles (the length of a row) minus $N/2$ (the number of columns).

\begin{figure}
    \centering
    \includegraphics[scale=0.6]{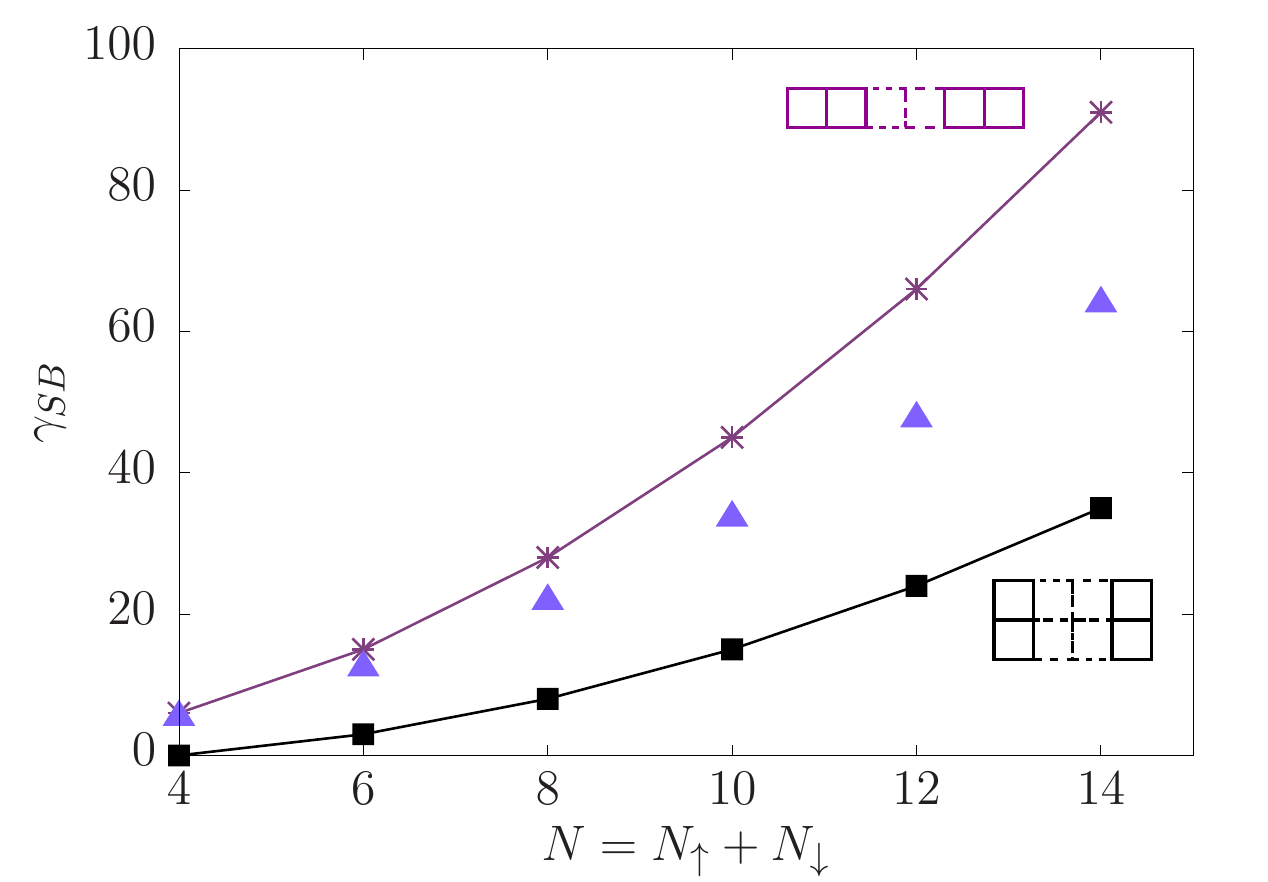}
    \caption{ $\gamma_{SB}$ as a function of $N$ (triangles) for the symmetry breaking ground-state. The stars and the boxes represent the eigenvalues $\gamma^{(2)}_S$ and $\gamma^{(2)}_A$ respectively. The lines are guide to the eye.}
    \label{fig-cas}
\end{figure}

In Fig. \ref{fig-cas}, we plot $\gamma_{SB}=\langle \Psi^{SB}|\Gamma^{(2)}|\Psi^{SB}\rangle$ as a function of $N$ and we compare it with $\gamma^{(2)}_S$
and $\gamma^{(2)}_A$. We observe that, by increasing $N$,  $\gamma_{SB}$ moves away from $\gamma^{(2)}_S$ to position itself halfway between $\gamma^{(2)}_S$ and $\gamma^{(2)}_A$. We have checked that the corresponding symmetry breaking ground state ($\vec{a}_P^{SB}$) does not correspond to any well-defined symmetry. The explicit calculation for the case of $N=4$ bosons is given in Appendix \ref{app-B}. The formal demonstration of the symmetry breaking is given in Appendix \ref{sec-gianni}.

\section{Correlation analysis}
\label{sec-corr}
The previous analysis allowed us to demonstrate that the precise protocol used in an experiment to set particle interactions to very large value has strong consequences on the symmetry properties of the ground state. However, the exchange symmetry, or the expectation of the two cycle class sum operators are not accessible experimentally. We therefore now look for a routinely-measured physical observable that would keep trace of the non-trivial symmetry of the ground state. The simplest one that strongly depends on the symmetry of the wave function is the momentum distribution, obtained from  %or in other words, 
the Fourier transform of the 
%single-particle 
one-body
density matrix. This statement is, for instance, obvious for non-interacting bosons and fermions which have completely different momentum distributions (Fermi-Dirac step function for fermions and Bose-Einstein distribution for bosons) but can be generalized to interacting mixtures with non trivial symmetries under exchange of particles \cite{Decamp2016-2,Decamp2017}.   
Starting from the many-body wave function, the one- body density matrix 
is obtained as follows: 
\begin{equation}
\rho_1(x,y)=N\int dx_2,... dx_N \Psi^* (x, x_2,...x_N) \Psi(y, x_2,....x_N),
\end{equation}
that, for a multi-component system, can be written \cite{Deuretzbacher2016}
\begin{equation}
    \rho_1(x,y)=\sum_\sigma N_\sigma \rho_{1,\sigma}(x,y)
    \end{equation}
    where
    \begin{equation}
    \rho_{1,\sigma}(x,y)=\sum_{i,j=1}^N c^{(i,j)}_{\sigma}\rho^{(i,j)}(x,y).
    \label{g_1}
\end{equation}
This representation is very useful as it separates spin and orbital correlations. Indeed, the term in Eq. (\ref{g_1}) 
\begin{widetext}
\begin{equation}
 \rho^{(i,j)}(x,y)   =\theta(x,y)N!\int_{x_1<\dots<x_{i-1}<x<x_{i+1}<\dots<x_j<y<x_{j+1}<\dots<x_N} 
 \!\!\!\!\!\!\!\!\!\!\!\!\!\!\!\!\!\!\!\!\!\!\!\!\!\!\!\!\!\!\!\!\!\!\!\!\!\!\!\!\!\!\!\!\!\!\!\!\!\!
 \!\!\!\!\! \!\!\!\!\!\!\!\!\!\!\!\!\!\!\!\!\!\!\!\!\!\!\!\! \!\!\!\!\!\!\!\!\!\!\!\!\!\!\!\!\!\!\!\!\!\!\!\!
 dx_1\dots dx_{i-1}dx_{i+1}\dots dx_{j-1}dx_{j+1}\dots dx_{N}
 \ \ |\Psi_S^*(x_1,\dots,x_{i-1},x,x_{i+1},\dots,x_N)\Psi_S(x_1,\dots,x_{j-1},y,x_{j+1},\dots,x_N)|,
 \label{eq:rhoij}
\end{equation}
\end{widetext}
where $\theta(x,y)$ is equal to 1 if $x\leq y$ and $i\leq j$, and 0 otherwise \cite{Deuretzbacher2016},
gives the probability that the $i$-th and the $j$-th particles are in $x$ and in $y$ positions respectively. The $y<x$ part of the correlation function can be obtained using the symmetry relation $\rho^{(i,j)}(x,y)=\rho^{(j,i)}(y,x)$.  
The term
\begin{equation}
    c^{(i,j)}_\sigma=\frac{N!}{\prod_\sigma N_\sigma!}\sum_{k=1}^{(N-1)!}a_{i(\sigma)k}a_{j(\sigma)k}.
    \label{eq:cij}
\end{equation}
 is the spin correlation function that gives the probability that the $i$-th and $j$-th particle have the same spin $\sigma$.
The amplitudes $a_{i(\sigma)k}$'s in Eq. (\ref{eq:cij}) are now labelled with respect to the position of the $i$-th particle ($i=1,...,N$) with spin $\sigma$ and consider all the $k=1,\dots,(N-1)!$ permutations of the $N-1$ other particles.
Remarkably, both the $SU(2)$ and the SB systems have the same spatial correlation function $\rho^{(i,j)}(x,y)$.
The symmetry properties affect only the spin correlation function $c^{(i,j)}_\sigma$.

As we are focusing on the special case of a ring that is invariant for translation symmetry, in the following we will set  $\rho_1(x,y)=\rho_1(x-y)=\rho_1(t)$ and $\rho_{1,\sigma}(x,y)=\rho_{1,\sigma}(x-y)=\rho_{1,\sigma}(t)$.
From an experimental point of view, one has easily access the momentum distribution, which is given by 
\begin{equation}
    n(k)=\int_{-L/2}^{L/2} e^{-ikt} \rho_1(t) {\rm d}t.
\end{equation}

\begin{figure}
    \centering
    \includegraphics[scale=0.6]{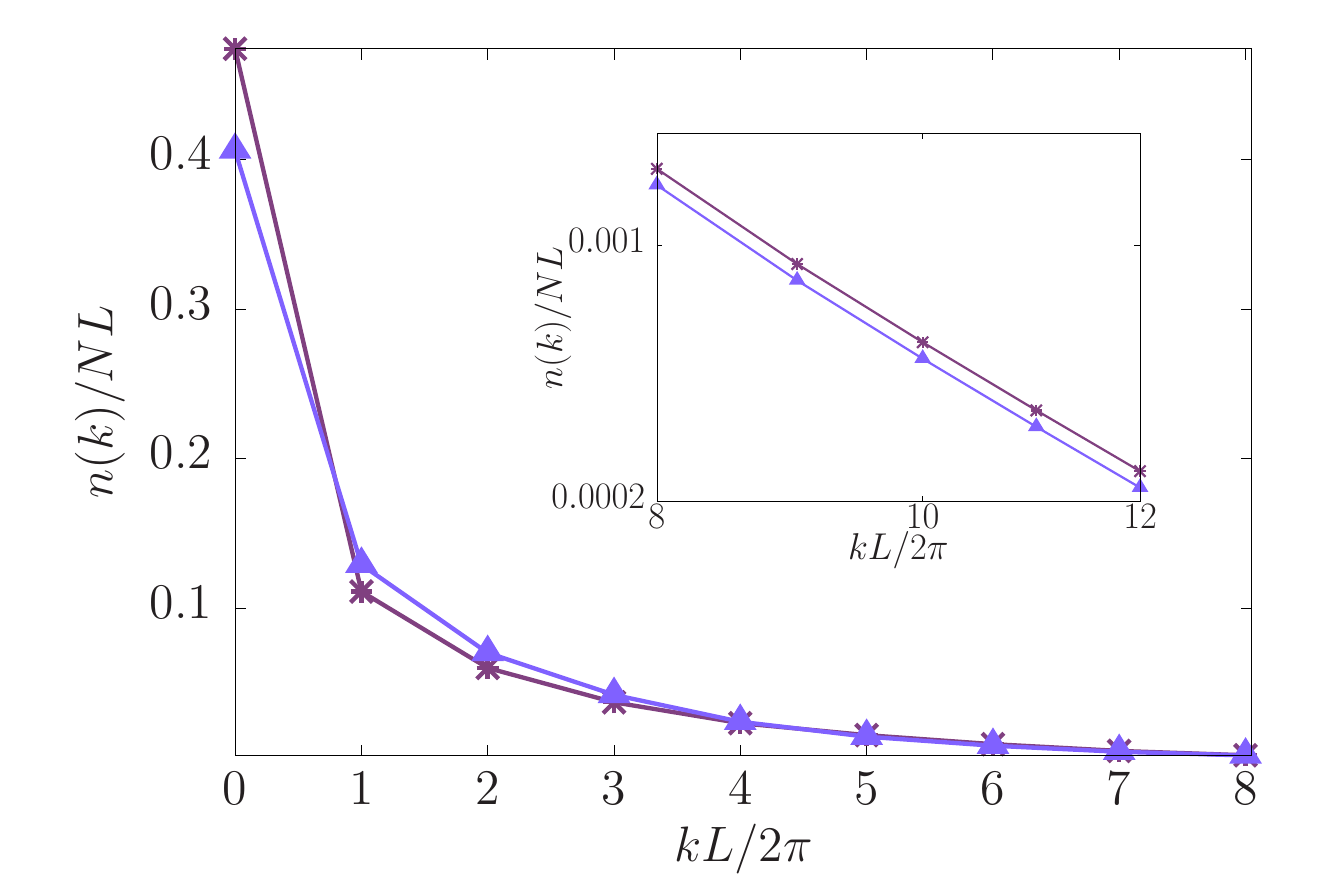}
    \caption{Normalized momentum distribution $n(k)/N$ in units of $1/L$ as a function of $kL/(2\pi)$ for a mixture of 4+4 bosons.
    The stars are the data for the $SU(2)$ mixture and the triangles for the SB system. The inset, in a log-log scale, is a zoom on the tails.
    The lines are guide to the eye.}
    \label{fig:mom}
\end{figure}

In Fig. \ref{fig:mom} we compare the momentum distribution for the SB mixture with that of
the ground-state for a $SU(2)$ mixture, the latter
coinciding with
the momentum distribution of a single component TG gas. We notice remarkable differences among the two both at small and at large momenta.
In particular, both the peak centered around $k=0$  and the tails at large $k$ (inset of Fig. \ref{fig:mom}) of the momentum distribution are larger for the $SU(2)$ mixture, while the one of the  SB mixture is higher at intermediate wavevectors.
Thus, in the following sections we will focus our study on $n_0=n(k=0)$ and $\lim_{k\rightarrow\infty}n(k)$ that provide information about large-distance and short-distance
correlations respectively. Moreover, in the ring geometry, $n_0$ coincides with the quasi-condensate fraction of the system.

\section{Large-distance correlations}
\label{sec-n0}
We now discuss in details the large-distance correlations which corresponds to small momenta in the momentum distribution. Specifically, we %even 
restrict this analysis to the zero-momentum occupation number which is an important measure of long-range coherence in quantum systems. For the balanced mixture discussed in this manuscript $\rho_{1,\sigma}(x,y)$ is independent of $\sigma$, so $n_0$ is given by
\begin{equation}
   n_0=\int^{L/2}_{-L/2}dt~\rho_1(t)
    =2\sum^{N/2}_{j=1}c_{1j}R^{j},
\end{equation}
where we have defined $R^{j}=\int^{L/2}_{-L/2}dt~\rho^{(1,j)}(t)$ and we have used that $c_{1,j}=c_{1,(N-j+1)}$.
We are mostly interested in the asymptotic behaviour at large number of particles that we approach by increasing the number of particles up to $N=14$. The results of our exact calculations are actually well approximated by a simple fitting function $R^j$ at large number of particles,
\begin{equation}
R^j\underset{N\rightarrow\infty}{\simeq}\dfrac{3}{4}\dfrac{1}{\sqrt{2j-1}}.
\end{equation}
For the $SU(2)$ case, $c_{1,j}=1$,  $\forall j$. This
implies that the ground state of the $SU(2)$ system coincides with that of a 
TG gas with a single 
spin component. Indeed if there was only one spin component, the spin correlation function would be maximum $\forall j$.
The resulting approximated expression for
the zero-momentum occupation, in the limit $N\gg 1$, reads
\begin{equation}
   n_0^{SU}(N\gg 1)\simeq  \dfrac{3}{2}\sum^{N/2}_{j=1}\dfrac{1}{\sqrt{2j-1}}.
   \label{no-su}
\end{equation}
This approximation Eq.(\ref{no-su}) provides the correct leading term of the function $n_0^{SU}(N)$ given in \cite{Forrester03} for a single component TG gas ,
\begin{equation}
   n_0^{SU}(N)=1.54\sqrt{N}-0.58+\dfrac{0.03}{\sqrt{N}}. 
   \label{forrester}
\end{equation}
For the SB case, the $c_{1j}$'s depend on $N$ for small values of $N$ but they seem to converge rapidly to a well defined value $c_{1j}$ for any $j$ (see Table \ref{tab:my_label}).
\begin{table}[t]
    \centering
    \begin{tabular}{|c|c|c|c|c|c|c|c|}
    \hline
       $N/2$  & $c_{12}$ & $c_{13}$ &  $c_{14}$ & $c_{15}$ & $c_{16}$& $c_{17}$& $c_{18}$\\
       \hline
        2& 0.833 &&&&&&\\
        \hline
        3& 0.811 & 0.769 &&&&&\\
        \hline
        4& 0.804 & 0.750 & 0.721 &&&&\\
        \hline
        5& 0.801 & 0.742 & 0.702 & 0.687 &&&\\
        \hline
        6& 0.799 & 0.737 & 0.692 & 0.671 & 0.660 &&\\
        \hline
        7& 0.798 & 0.735 & 0.687 & 0.662 & 0.645 & 0.638 &\\
        \hline
        8& 0.797 & 0.733 & 0.683 & 0.656 & 0.636 & 0.625 & 0.619\\
        \hline
    \end{tabular}
    \caption{Behaviour of the absolute value of the coefficients $c_{1j}$ as functions of $N$ for the case of breaking symmetry.}
    \label{tab:my_label}
\end{table}
Breaking the $SU(2)$ symmetry makes the two spin states distinguishable. Thus, we expect that, at large $j$, there are no more correlations between the first spin and 
the $j$-th one, so that the probability $c_{1j}$ to have the same spin state
has to tend to $1/2$.
Indeed the $c_{1j}$'s can be fitted with the function
\begin{equation}
    f_{1j}=\left(\dfrac{1}{2}+\dfrac{1}{2}e^{-b(j-1)^a}\right),
 \label{fit-cij}  
\end{equation}
$a$ and $b$ being positive and slightly depending on $N$.

\begin{figure}
     \centering
     \includegraphics[scale=0.6]{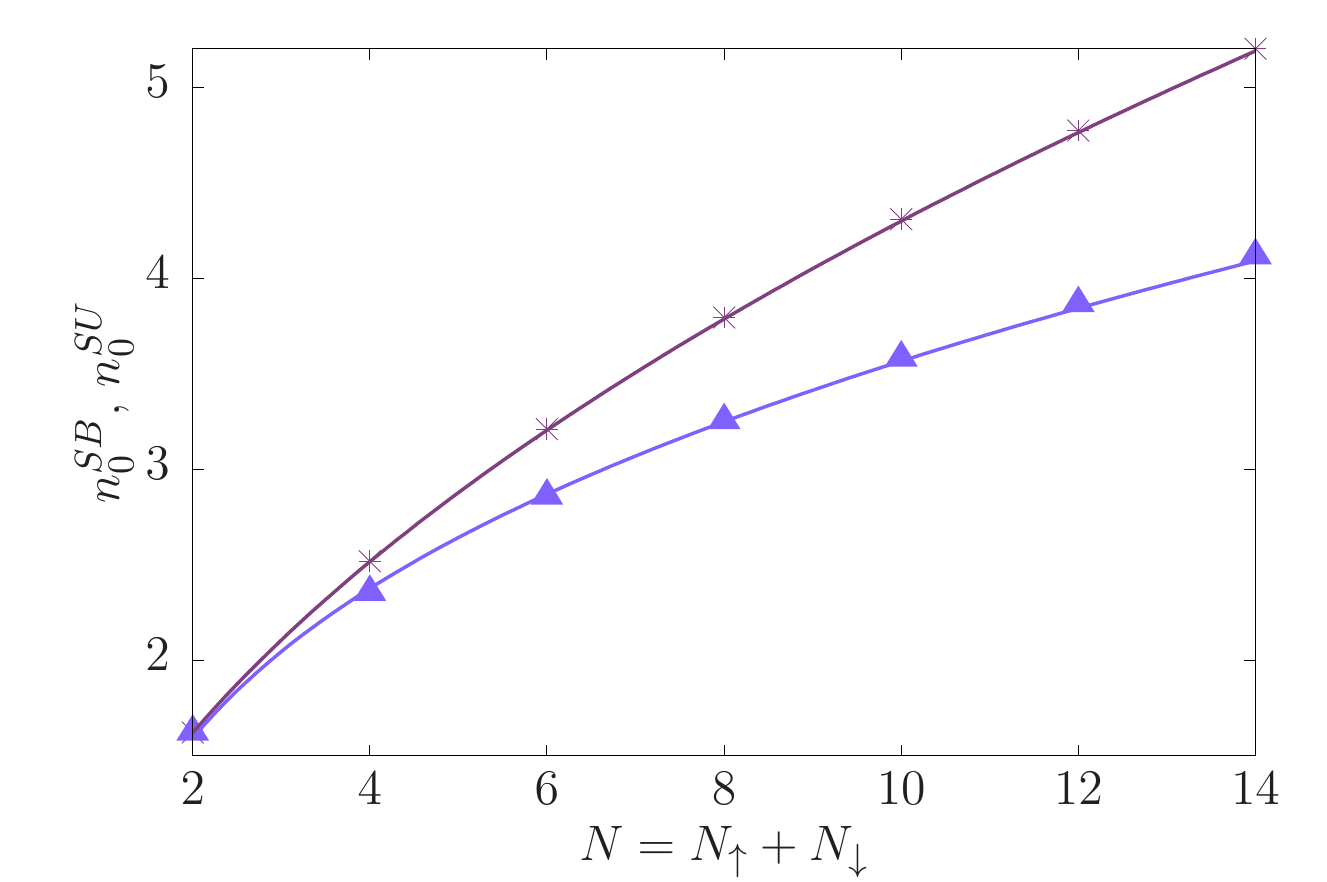}
     \caption{The zero-momentum occupation numbers $n_0^{SU}$ (stars) and $n_0^{SB}$ (triangles), for a balanced mixture, as a function of the total number of particles $N$. The exact results (points) are compared with the approximated function (lines) given respectively in Eq. (\ref{forrester}) and Eq. (\ref{gianni}).}
     \label{fig_no}
 \end{figure}

The exponential decay part of Eq.~(\ref{fit-cij}) does not contribute in the thermodynamic limit, so that 
 \begin{equation}
     \lim_{N\rightarrow\infty}\dfrac{n_0^{SB}}{n_0^{SU}}=\dfrac{1}{2}.
 \end{equation}
 In Fig. \ref{fig_no} we plot the exact results for $n_0^{SU}$ and $n_0^{SB}$,
 together with the analytical approximated expression for $n_0^{SU}(N)$  given in Eq. (\ref{forrester}) and that for the symmetry-breaking case, 
 \begin{equation}
    n_0^{SB}(N)=0.77\sqrt{N}+1.64-\dfrac{1.61}{\sqrt{N}}.
    \label{gianni}
 \end{equation}
 Eq.~(\ref{gianni}) has been obtained by fitting the data obtained by the exact calculation and by fixing the first coefficient to 0.77 (half the first coefficient of Eq.~\ref{forrester})).
 
 Breaking the $SU(2)$ symmetry has therefore the tendency to destroy long range coherence. For our particular model, the zero-momentum occupation number is reduced by a factor of two. This macroscopic consequence of a microscopic symmetry property is a central result of this paper as it constitutes an experimental smoking gun of $SU(2)$ symmetry breaking.
 
\section{Short-distance correlations: the Tan's contact}
\label{sec-tan}
We now proceed with the discussion of short-distance correlations. This time, they are observable in the tails of the momentum distribution. For a  system
with zero-range interactions, the momentum distribution decays as $k^{-4}$. The prefactor
 $\mathcal{C}=\lim_{k\rightarrow\infty}n(k)k^4$ is the so-called Tan's contact \cite{Tan2008a}.
 This observable is proportional to the cusps in the systems, namely to the symmetric exchanges between
 particles \cite{Decamp2016-2,Decamp2017}.
 In this section we will focus on the modification of the Tan's contact due to symmetry breaking.

For the $SU(2)$-symmetric system, the Tan's contact is proportional to the energy slope $K^{SU}$,
\begin{equation}
     \mathcal{C}^{SU}=\frac{2m^2}{\hbar^4}K^{SU}.
     \end{equation}
 
We see from the cusp conditions (\ref{cuspC}) and (\ref{cuspC2}) that in the ring geometry for the $SU(2)$ case, there are $N$ cusps, and each cusp brings a contribution
that is proportional to twice $\alpha^{(N)}$, so that
$K^{SU}=2N\alpha^{(N)}\hbar^4/m^2$ and thus $\mathcal{C}^{SU}=4N\alpha^{(N)}$.
 
For the SB case, $K^{SB}_{\uparrow\downarrow}$ takes into account only the inter-component
contribution, as our starting point in the energy calculation is a two-component TG
gas whose intra-species interaction strength is set to infinity from the beginning.
However, the Tan's contact is related to both the intra- and inter-component contributions
$[\partial_{1/g_{\sigma,\sigma}}E]_{g_{\sigma,\sigma}\rightarrow \infty}$ and $[\partial_{1/g_{\sigma,\sigma'}}E]_{g_{\sigma,\sigma'}\rightarrow \infty}$, the first term counting
the cusps for 
exchange of identical bosons, and the second giving the cusps for  
exchange of bosons
with different spins.
Specifically, in the SB case the contact 
is given by (see App.~\ref{lastapp} for derivation)
\begin{equation}
     \mathcal{C}^{SB}=\frac{2m^2}{\hbar^4}\left[(\Vec{a}_P^{SB})^t  V^{SU}\Vec{a}_P^{SB}\right]
     \label{lanna}
\end{equation}
with $\Vec{a}_P^{SB}$ is the eigenvector of $V^{SB}$ corresponding to its largest eigenvalue.

\begin{figure}
    \centering
    \includegraphics[scale=0.6]{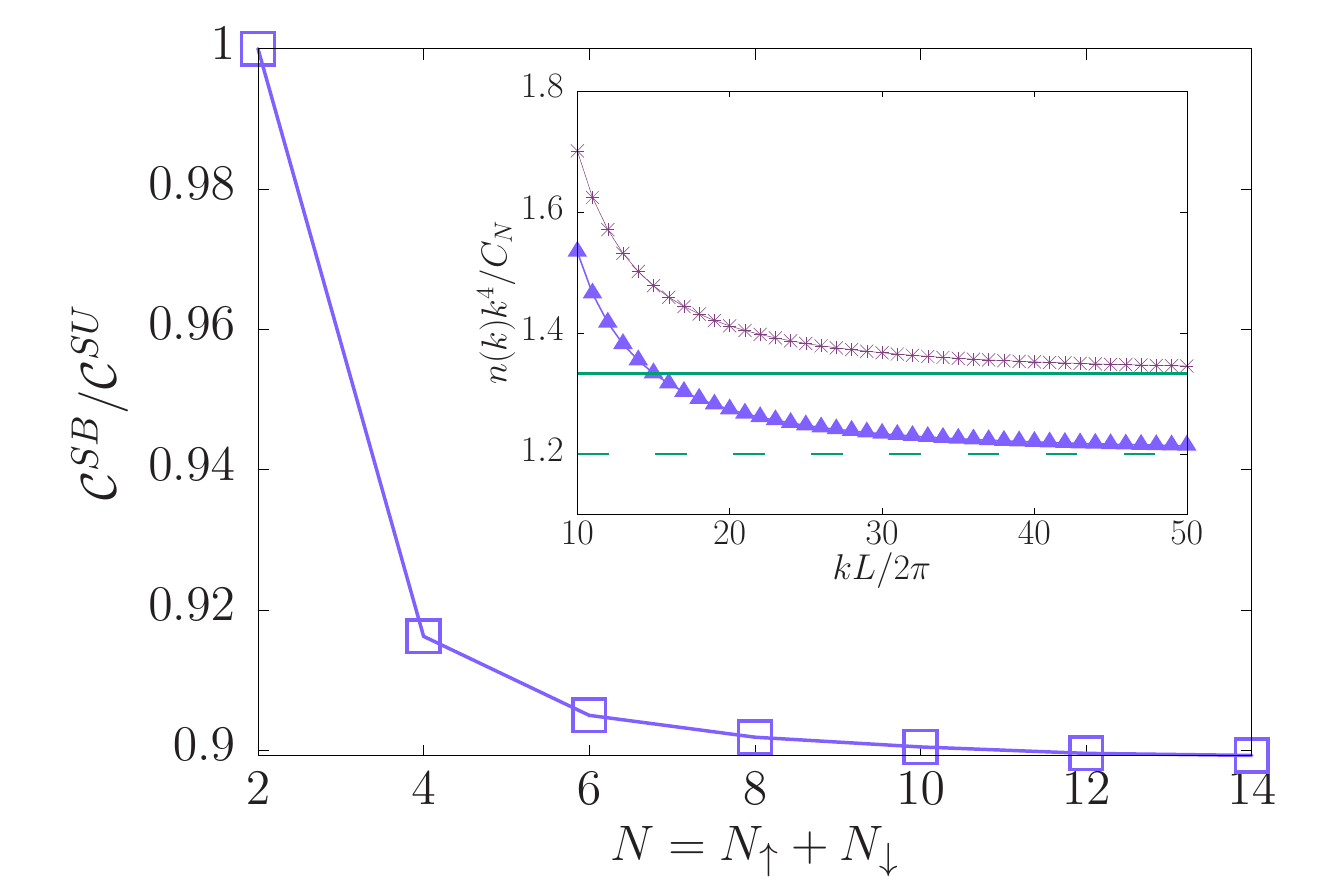}
    \caption{The ratio $\mathcal{C}^{SB}/\mathcal{C}^{SU}$ as a function of $N=N_\uparrow+N_\downarrow$ for balanced mixtures (the line is a guide to the eye).
    In the inset we show $N(k)k^4$, in units of  $C_N=N^2(N^2-1)/L^3$, as a function of $kL/(2\pi)$ for the case of a $SU(2)$ mixture (stars) and a SB one (triangles) of $N=4+4$ bosons. The horizontal lines indicates the values of $\mathcal{C}^{SU}/C_N$ (continuous line) and $\mathcal{C}^{SB}/C_N$ (dashed line).}
    \label{fig_ratio}
\end{figure}

In Fig. \ref{fig_ratio} we plot the ratio $\mathcal{C}^{SB}/\mathcal{C}^{SU}$
as a function of $N$. We observe
that $\mathcal{C}^{SB}/\mathcal{C}^{SU}$ converges very rapidly to $\sim 0.9$. Thus for $N>2$, the contact is lower for the SB case than for the $SU(2)$ mixture. As reported for other multicomponent mixtures
\cite{Decamp2016-2}, the 
reduction
of the symmetry also manifests itself in the lowering
of the contact   in this case. The fact that the change  
is relatively small is due to the fact that each 
component of the mixture is bosonic and, then, several cusps are still present in the SB case.

\section{Concluding remarks}
\label{sec-concl}
In this paper we have presented a model of a boson-boson mixture where exchange symmetry is broken and obtained its solutions at large inter-particle interaction. Before summing up our conclusions, we would like to mention that the solution of such a model can also be obtained, for any strength of the inter-particle interaction,  
by means of the Bethe Ansatz solution for the Yang-Gaudin Hamiltonian\cite{Gaudin67}.
Indeed one can write, in each coordinate sector $Q$ such that
$x_{Q(1,\uparrow)}<\dots<x_{Q(N,\downarrow)}$,
\begin{equation}
    \Psi^{SB}_Q(x_1,\dots,x_N)=\prod_{i,j}\prod_{\sigma=\uparrow,\downarrow} {\rm sign}(x_{Q(i,\sigma)}-x_{Q(j,\sigma)})\Psi^{YG}_Q(x_1,\dots,x_N)
\end{equation}
where the function $\Psi^{YG}_Q$ is
the Bethe wavefunction for the $SU(2)$ Fermi gas 
in the coordinate sector $Q$. The great advantage of our method, that is exact up to the order
$1/g_{\uparrow\downarrow}$, is the ease with which one can access the one-body correlation function,
allowing a deep understanding of spatial and spin correlations.
Another
important  advantage of the method outlined in this work is that it can be applied to any trapping potential. As soon as one knows the single-particle orbitals, such as for the case of a harmonic potential or a box trap, it is possible to write the exact solution for the many-body wavefunction for the symmetry-breaking case too. 

In particular, in this work, we have shown that different spin states with different symmetries can be obtained by
varying the protocol used in order to achieve the strong-repulsive limit. The symmetry breaking  induced by the difference between the intra- and inter-specie interaction strengths affects both short- and large-distance correlations, but
the effect on the large-distance correlations is more dramatic. Indeed, at large number of particles we observe  a depletion by a factor two of the zero-momentum occupation number, which is a signature of a lack of spin correlation at large distance. This means that the zero-momentum occupation number is a very sensitive observable for detecting symmetry breaking. 

Our work provides
a guide for
the studies of the correlation properties of $SU(\kappa)$ mixtures in the strongly interacting regime, highlighting the importance of the protocol chosen to reach such a regime.

\acknowledgements
 G. A.-D. acknowledges Manon Ballu and Martial Morisse for fruitful discussions, and all authors aknowledge H\'el\`ene Perrin for important comments on the symmetry analysis.  We acknowledge funding from the ANR-21-CE47-0009 Quantum-SOPHA project.
 \appendix
 \section{The matrices $V^{SU}$ and $V^{SB}$ }
 \label{app-A}
 Here we will give the explicit example for the calculation of $V^{SU}$ and $V^{SB}$ for the case of a balanced mixture with $N=4$ bosons.
We consider the snippet basis $\{$$\uparrow\uparrow\downarrow\downarrow$,
$\uparrow\downarrow\uparrow\downarrow$, $\uparrow\downarrow\downarrow\uparrow$,
$\downarrow\uparrow\uparrow\downarrow$,
$\downarrow\uparrow\downarrow\uparrow$, 
$\downarrow\downarrow\uparrow\uparrow$
$\}$. 
For the $SU(2)$ mixture in a ring geometry, the $V^{SU}$ matrix 
reads

 \begin{equation}
V^{SU}=\frac{\hbar^4}{m^2}\alpha^{(N)}\left(\begin{array}{cccccc}
 6& 1&0&0&1&0  \\
 1&4&1&1&0&1\\
 0& 1&6&0&1&0\\
 0& 1&0&6&1&0\\
 1&0& 1& 1&4& 1\\
 0& 1&0&0&1&6
\end{array}
\right),
\label{matr-vsu}
\end{equation}
whose largest eigenvalue is $8\hbar^4\alpha^{(N)}/m^2$ with corresponding eigenvector 
$\Vec{a}_P^{SU}=\frac{1}{\sqrt{6}}(1,1,1,1,1,1)$.

For the SB mixture, the $V^{SB}$ matrix 
reads

 \begin{equation}
V^{SB}=\frac{\hbar^4}{m^2}\alpha^{(N)}\left(\begin{array}{cccccc}
  2& 1&0&0&1&0  \\
 1&4&1&1&0&1\\
 0& 1&2&0&1&0\\
 0& 1&0&2&1&0\\
 1&0& 1& 1&4& 1\\
 0& 1&0&0&1&2
\end{array}
\right).
\label{matr-vsb}
\end{equation}
The largest eigenvalue is $6\hbar^4\alpha^{(N)}/m^2$ and its corresponding eigenvector reads
$\Vec{a}_P^{SB}=\frac{1}{2\sqrt{3}}(1,2,1,1,2,1)$.

It is worth making the case $N=2$ explicitly. Indeed, because of the periodic boundary conditions, the $\delta(x_1-x_2)$
contributes twice both for the diagonal terms and the off-diagonals ones. Thus, on the snippet basis $\{\uparrow\downarrow,\downarrow\uparrow\}$, one has the matrix
\begin{equation}
V^{SU}=V^{SB}=\frac{\hbar^4}{m^2}\alpha^{(N)}\left(\begin{array}{cc}
2& 2\\
2& 2
\end{array}
\right),
\end{equation}
whose largest eigenvalue is $4\alpha^{(N)}\hbar^4/m^2$, in agreement with $K^{SU}=2N\alpha^{(N)}\hbar^4/m^2$.

\section{Mapping on the $XXX$ spin-chain model for $SU(2)$ mixtures}

\label{app-xxx}
In the strong-interacting limit, in the case of $SU(2)$ bosons or fermions, the Hamiltonian (\ref{ham}) can be mapped into a spin-chain model. Indeed, at the order $1/g$
one can write \cite{Deuretzbacher2014}
\begin{equation}
    \hat{H}-\mathbb{1}E_{g\rightarrow\infty}=-V^{SU}_{B,F}/g= -NJ\mathbb{1}\mp J\sum_{j=1}^{N}\hat P_{j,j+1}
    \label{SpinCkappa}
\end{equation}
where $J=\alpha^{(N)}/g$, the - (+) sign applies to bosons (fermions).
Since the permutation operator $\hat P_{j,j+1}$ can be written as a function of
product of Pauli matrices $\hat P_{j,j'}=(\vec \sigma^{(j)}\vec\sigma^{(j')}+\mathbb{1})/2$ acting on site $j$ and $j'$,
it is straightforward to show that it is possible to map (\ref{SpinCkappa}) on a Heisenberg XXX chain model, both for bosons and fermions: a ferromagnetic one for $SU(2)$ bosons,
\begin{equation}
-\frac{V^{SU}_{B}}{g} =-2J \sum_{j=1}^N\vec{S}^{(j)}\vec{S}^{(j+1)}-\dfrac{3}{2}NJ\mathbb{1}
\end{equation}
and an antiferromagnetic one for $SU(2)$ fermions,
\begin{equation}
-\frac{V^{SU}_{F}}{g} =2J \sum_{j=1}^N\vec{S}^{(j)}\vec{S}^{(j+1)}-\dfrac{1}{2}NJ\mathbb{1},
\end{equation}
where $\vec{S}=\vec{\sigma}/2$ are the spin operators.

For the SB case the Hamiltonian can be written
\begin{equation}
\begin{split}
    \hat{H}-\mathbb{1}E_{g\rightarrow\infty}&=-V^{SB}/g\\ &= -NJ\mathbb{1}- J\sum_{j=1}^{N}\hat P_{j,j+1} +2 J\sum_{j=1}^{N}
    |s\rangle\langle s|\hat P_{j,j+1} |s\rangle\langle s|
    \label{SpinCkappa2}
    \end{split}
\end{equation}
where $|s\rangle\langle s|$ is the projector on the snippet basis, so that the last term applies only on diagonal elements.
From this writing, it is clear the origin of the SB: the term $- J\sum_{j=1}^{N}\hat P_{j,j+1}$
is the bosonic one, while the term $+2 J\sum_{j=1}^{N}
    |s\rangle\langle s|\hat P_{j,j+1} |s\rangle\langle s|$ is at the origin of a partial fermionization acting only partially on the system (on the diagonal terms).
One can show that
\begin{equation}
 2 J\sum_{j=1}^{N}
    |s\rangle\langle s|\hat P_{j,j+1} |s\rangle\langle s|=J\sum_{j=1}^{N}(\mathbb{1}+4S_z^{(j)}S_z^{(j+1)})   
\end{equation}
Thus we get a XXZ Heisenberg chain Hamiltonian:
\begin{equation}
-\frac{V^{SB}}{g} =-2J \sum_{j=1}^N(S_x^{(j)}S_x^{(j+1)}+S_y^{(j)}S_y^{(j+1)}-S_z^{(j)}S_z^{(j+1)})-\dfrac{1}{2}NJ\mathbb{1}.
\end{equation}
Remark that such a XXZ Hamiltonian can me mapped on a XXX one with an opposite sign of $J$ by applying the unitary transformation $U=\prod_{\ell=even}2S_z^{(\ell)}$ \cite{Takahashi,Volosniev2015}.
Such operator {\it does not preserve the symmetry} (does not commute with the operator $\Gamma^{(2)}$) and its action is equivalent to map TG bosons on non-interacting fermions and vice-versa.
On our snippet basis,
\begin{equation}
    U=\left(\begin{array}{cccccc}
 -1& 0&0&0&0&0  \\
 0&1&0&0&0&0\\
 0& 0&-1&0&0&0\\
 0& 0&0&-1&0&0\\
 0&0& 0& 0&1& 0\\
 0& 0&0&0&0&-1
\end{array}
\right).
\end{equation}

\section{The matrix $\Gamma^{(2)}$}
\label{app-B}
For the case of a balanced mixture of $N=4$ bosons, the $\Gamma^{(2)}$ matrix can be written in the snippet basis (taking into account the initial Ansatz for the many-body wavefunction (\ref{vol})) as
\begin{equation}
\Gamma^{(2)}=\left(\begin{array}{cccccc}
 2& 1&1&1&1&0  \\
 1&2&1&1&0&1\\
 1& 1&2&0&1&1\\
 1& 1&0&2&1&1\\
 1&0& 1& 1&2& 1\\
 0& 1&1&1&1&2
\end{array}
\right)
\label{matr-gamma2}
\end{equation}
which can be diagonalized. This yields three representations of dimension 1,3 and 2 with eigenvalues $\gamma_2=6,2,0$ corresponding to the diagrams $\tiny\yng(4)$, $\tiny\yng(3,1)$ and $\tiny\yng(2,2)$. The eigenstate corresponding the irreductible representation of dimension one is 
$\Vec{\upsilon_6}=\frac{1}{\sqrt{6}}(1,1,1,1,1,1)$ which is identical to the ground state of the $SU(2)$ model. 
The other eigenvectors are $\vec{\upsilon}_{2_1}=\frac{1}{\sqrt{2}}(-1,0,0,0,0,1)$, $\vec{\upsilon}_{2_2}=\frac{1}{\sqrt{2}}(0,-1,0,0,1,0)$, $\vec{\upsilon}_{2_3}=\frac{1}{\sqrt{2}}(0,0,-1,1,0,0)$, $\vec{\upsilon}_{0_1}=\frac{1}{2}(1,0,-1,-1,0,1)$ and $\vec{\upsilon}_{0_2}=\frac{1}{2\sqrt{3}}(1,-2,1,1,-2,1)$.
The ground state of the system with broken $SU(2)$ symmetry will be a linear superposition of states with different symmetries. In this precise case we obtain that $\Vec{a}_P^{SB}=\frac{2\sqrt{2}}{3}\vec{\upsilon}_6+\frac{1}{3}\vec{\upsilon}_{0_2}$, namely the symmetries involved are  mainly $\tiny\yng(4)$ ($\frac{8}{9}$) but also $\tiny\yng(2,2)$ ($\frac{1}{9}$) .

\section{Demonstration of the $SU(\kappa)$ symmetry breaking}
\label{sec-gianni}
For the case of $N=4$ one can easily calculate the commutators  $[V^{SU},\Gamma^{(2)}]$ and $[V^{SB},\Gamma^{(2)}]$ using the explicit form of the matrices (\ref{matr-vsu}), (\ref{matr-vsb}) and (\ref{matr-gamma2}).
The first is zero, legitimating the use of the Young tableaux for the identification of the symmetries of the eigenstates, while the second is different from zero, that is a proof of SB.

In this appendix we  generalise our demonstration to the case of arbitrary $N$ particles, for a homogeneous system or in the case of an inhomogeneous trapping potential
(including a site dependence on the exchange $J$ terms, $J\rightarrow J_j$), and extending
the discussion to any $SU(\kappa)$ mixture. Indeed the introduction of other spin components only affects the definition of the snippet basis.

Let us start by writing the matrices $\Gamma^{(2)}$ and $V^{SU}$ as function of the permutation operators.
Using the (\ref{SpinCkappa}) form of $V^{SU}/g$, we see that the commutator $[V^{SU},\Gamma^{(2)}]$ reduced to $\sim [\sum_{i}J_i\hat{P}_{i,i+1},\sum_{i'<j'}\hat{P}_{i',j'}]=\frac{1}{2}[\sum_{i}J_i\hat{P}_{i,i+1},\sum_{i',j'}\hat{P}_{i',j'}]$.
Then one can write the permutations operators in the second quantization framework as following \cite{Auerbach}

\begin{equation}
    \hat{P}_{i,j}=\sum_{\mu,\nu}F_\mu^\nu(i)F_{\nu}^{\mu}(j),
\end{equation}

where $F_\mu^\nu(i)=a^\dagger_{i,\mu}a_{i,\nu}$, and $a$ and $a^\dagger$ are usual annihilation and creation operators (fermionic or bosonic),  $\mu$ and $\nu$ are the spin-$s$ projection indices going from $1$ to $2s+1$, and $i$ and $j$ are the sites indices. It is important to notice that the $F_\mu^\nu(i)$'s are the generators of the $SU(\kappa)$ group, satisfying the commutation relation of the $SU(\kappa)$ Lie algebra
\begin{equation}
    \left[F_\mu^\nu(i),F_{\nu'}^{\mu'}(j)\right]=\delta_i^j\left(\delta^{\nu}_{\mu'}F_{\mu}^{\nu'}(i)-\delta_{\mu}^{\nu'}F_{\mu'}^{\nu}(j)\right).
    \label{comFmunu}
\end{equation}

By using the commutation relation (\ref{comFmunu}), one find that $[\sum_{i}J_i P_{i,j},\sum_{i',j'} P_{i',j'}]=0$ for any $j$, and thus also for $j=i+1$.
Starting with the (\ref{SpinCkappa2}) form of $V^{SB}$ one can find that $[V^{SB}/g,\Gamma^{(2)}]=[4\sum_jJ_jS_z^{(j)}S_z^{(j+1)},2\sum_{n,j'}\vec{S}^{(j')}\vec{S}^{(j'+n)}]$. Again, using the commutation relation of spin matrices, it is straightfoward to obtain that

\begin{widetext}
\begin{equation}
 \left[V^{SB}/g,\Gamma^{(2)}\right]=8
    \sum_{j,n}J_j\left\{\left(S_+^{(j)}S_-^{(j+n)}-S_-^{(j)}S_+^{(j+n)}\right)\left(S_z^{(j+n+1)}-S_z^{(j+1)}\right)+\left(S_z^{(j+n-1)}-S_z^{(j-1)}\right)\left(S_+^{(j)}S_-^{(j+n)}-S_-^{(j)}S_+^{(j+n)}\right)\right\},
\end{equation}
\end{widetext}

that doesn't vanish regardless the type of mixture ($\forall \kappa$) and/or the number of particles ($N>2$). 

\section{demonstration of Eq. (\ref{lanna})}
\label{lastapp}
Let us consider the Hamiltonian (\ref{ham})
\begin{equation}
    \hat H=H_{kin}+ \sum_{\sigma=\uparrow,\downarrow}\hat H_{int, \sigma\sigma}+H_{int, \uparrow\downarrow},
\end{equation}
where we have defined $H_{kin}=\sum_{\sigma=\uparrow,\downarrow}\sum_i^{N_\sigma}
-\frac{\hbar^2}{2m}\frac{\partial^2}{\partial x_{i,\sigma}^2}$,
$H_{int, \sigma\sigma}=
g_{\sigma\sigma}\sum_i^{N_\sigma}\sum_{j>i}^{N_{\sigma}} \delta(x_{i,\sigma}-x_{j,\sigma})$, and $H_{int,\uparrow\downarrow}=
g_{\uparrow\downarrow}\sum_{i}^{N_\uparrow}\sum_{j}^{N_\downarrow}\delta(x_{i,\uparrow}-x_{j,\downarrow})$.

The first step is to make the Fourier Transform of the Schr\"odinger equation $\hat H\Psi=E\Psi$ with respect, for instance, to $x_1$.
Let $\sigma$ be the spin of such particle.
In the large-momentum limit, using that $\lim_{k\rightarrow\infty}\Psi(k,x_2,\dots,x_N)=0$, one gets
\begin{equation}
\begin{split}
&\lim_{k\rightarrow\infty} \dfrac{\hbar^2k^2}{2m}\Psi(k,x_2,\dots,x_N)=g_{\uparrow\downarrow}\!\!\!\! \sum_{j,\sigma'\neq\sigma}\!\!\!\Psi(x_{j,\sigma'}, \dots,x_{j,\sigma'},\dots)e^{-ikx_{j,\sigma'}}\\&+\sum_{\sigma=\uparrow,\downarrow}g_{\sigma\sigma} \sum_{j,\sigma}\Psi(x_{j,\sigma}, \dots,x_{j,\sigma},\dots)e^{-ikx_{j,\sigma}}.
\end{split}
\end{equation}
By multiplying by the complex conjugate, one obtains the following asymptotic behaviour of the total momentum distribution $n(k)$:
\begin{equation}
\begin{split}
   & \lim_{k\rightarrow\infty} k^4 n(k)= \\&\dfrac{2 m^2}{\hbar^4} \left(  g_{\uparrow\downarrow}\langle\Psi|\hat H_{int, \uparrow\downarrow}|\Psi\rangle+ 
  \sum_{\sigma=\uparrow,\downarrow}g_{\sigma\sigma} \langle\Psi|\hat H_{int, \sigma\sigma}|\Psi\rangle\right)
    \end{split}
    \label{finire}
\end{equation}
By applying the Hellmann–Feynman theorem, it is straightforward to show that Eq.~(\ref{finire}), with $\Psi$ the symmetry-breaking many-body wavefunction, gives Eq.~(\ref{lanna}) in the Tonks-Girardeau limit.

\end{document}